\documentclass[oneside,a4paper,12pt,reqno]{amsart}
\usepackage{amsmath, amstext,amsbsy,amsopn,amsthm, amscd,amsfonts,amssymb}
\usepackage{graphicx}
\usepackage[colorlinks=true,citecolor=blue,linkcolor=blue]{hyperref}

\usepackage{times}
\usepackage{mathptmx}
\usepackage[round]{natbib}
\providecommand{\tabularnewline}{\\}

\date{}

\begin{document}
\global\long\def\dg{\mathbf{F}}
\global\long\def\dgcomp#1{F_{#1}}
\global\long\def\piola{\mathbf{P}}
\global\long\def\refbody{\Omega_{0}}
\global\long\def\refbnd{\partial\refbody}
\global\long\def\bnd{\partial\Omega}
\global\long\def\rcg{\mathbf{C}}
\global\long\def\lcg{\mathbf{b}}
\global\long\def\rcgcomp#1{C_{#1}}
\global\long\def\cronck#1{\delta_{#1}}
\global\long\def\lcgcomp#1{b_{#1}}
\global\long\def\deformation{\boldsymbol{\chi}}
\global\long\def\dgt{\dg^{\mathrm{T}}}
\global\long\def\idgcomp#1{F_{#1}^{-1}}
\global\long\def\velocity{\mathbf{v}}
\global\long\def\accel{\mathbf{a}}
\global\long\def\vg{\mathbf{l}}
\global\long\def\idg{\dg^{-1}}
\global\long\def\cauchycomp#1{\sigma_{#1}}
\global\long\def\idgt{\dg^{\mathrm{-T}}}
\global\long\def\cauchy{\boldsymbol{\sigma}}
\global\long\def\normal{\mathbf{n}}
\global\long\def\normall{\mathbf{N}}
\global\long\def\traction{\mathbf{t}}
\global\long\def\tractionl{\mathbf{t}_{L}}
\global\long\def\ed{\mathbf{d}}
\global\long\def\edcomp#1{d_{#1}}
\global\long\def\edl{\mathbf{D}}
\global\long\def\edlcomp#1{D_{#1}}
\global\long\def\ef{\mathbf{e}}
\global\long\def\efcomp#1{e_{#1}}
\global\long\def\efl{\mathbf{E}}
\global\long\def\freech{q_{e}}
\global\long\def\surfacech{w_{e}}
\global\long\def\outer#1{#1^{\star}}
\global\long\def\perm{\epsilon_{0}}
\global\long\def\matper{\epsilon}
\global\long\def\jump#1{\left[\left[#1\right]\right]}
\global\long\def\identity{\mathbf{I}}
\global\long\def\area{\mathrm{d}a}
\global\long\def\areal{\mathrm{d}A}
\global\long\def\refsys{\mathbf{X}}
\global\long\def\Grad{\nabla_{\refsys}}
\global\long\def\grad{\nabla}
\global\long\def\divg{\nabla\cdot}
\global\long\def\Div{\nabla_{\refsys}}
\global\long\def\derivative#1#2{\frac{\partial#1}{\partial#2}}
\global\long\def\aef{\Psi}
\global\long\def\dltendl{\edl\otimes\edl}
\global\long\def\tr#1{\mathrm{tr}\left(#1\right)}
\global\long\def\ii#1{I_{#1}}
\global\long\def\dh{\hat{D}}
\global\long\def\inc#1{\dot{#1}}
\global\long\def\sys{\mathbf{x}}
\global\long\def\curl{\nabla}
\global\long\def\Curl{\nabla_{\refsys}}
\global\long\def\piolaincpush{\boldsymbol{\Sigma}}
\global\long\def\piolaincpushcomp#1{\Sigma_{#1}}
\global\long\def\edlincpush{\check{\mathbf{d}}}
\global\long\def\edlincpushcomp#1{\check{d}_{#1}}
\global\long\def\efincpush{\check{\mathbf{e}}}
\global\long\def\efincpushcomp#1{\check{e}_{#1}}
\global\long\def\elaspush{\boldsymbol{\mathcal{C}}}
\global\long\def\elecpush{\boldsymbol{\mathcal{A}}}
\global\long\def\elaselecpush{\boldsymbol{\mathcal{B}}}
\global\long\def\disgrad{\mathbf{h}}
\global\long\def\disgradcomp#1{h_{#1}}
\global\long\def\trans#1{#1^{\mathrm{T}}}
\global\long\def\phase#1{#1^{\left(p\right)}}
\global\long\def\elecpushcomp#1{\mathcal{A}_{#1}}
\global\long\def\elaselecpushcomp#1{\mathcal{B}_{#1}}
\global\long\def\elaspushcomp#1{\mathcal{C}_{#1}}
\global\long\def\dnh{\aef_{DH}}
\global\long\def\dnhc{\mu\lambda^{2}}
\global\long\def\dnhcc{\frac{\mu}{\lambda^{2}}+\frac{1}{\matper}d_{2}^{2}}
\global\long\def\dnhb{\frac{1}{\matper}d_{2}}
\global\long\def\afreq{\omega}
\global\long\def\dispot{\phi}
\global\long\def\edpot{\varphi}
\global\long\def\newcofa{A_{1}^{+}}
\global\long\def\newcofaa{A_{5}^{+}}
\global\long\def\newcofb{B^{-}}
\global\long\def\newcofc{C^{+}}
\global\long\def\norma{\hat{A}_{1}^{+}}
\global\long\def\normb{\hat{B}_{1}^{+}}
\global\long\def\normc{\hat{C}_{1}^{+}}
\global\long\def\normaa{\hat{A}_{3}^{+}}
\global\long\def\kh{\hat{k}}
\global\long\def\afreqh{\hat{\afreq}}
\global\long\def\phasespeed{c}
\global\long\def\bulkspeed{c_{B}}
\global\long\def\speedh{\hat{c}}
\global\long\def\dhth{\dh_{th}}
\global\long\def\bulkspeedl{\bulkspeed_{\lambda}}
\global\long\def\khth{\hat{k}_{th}}
\global\long\def\p#1{#1^{\left(p\right)}}
\global\long\def\maxinccomp#1{\inc{\outer{\sigma}}_{#1}}
\global\long\def\maxcomp#1{\outer{\sigma}_{#1}}
\global\long\def\relper{\matper_{r}}
\global\long\def\sdh{\hat{d}}
\global\long\def\iee{\varphi}
\global\long\def\effectivemu{\tilde{\mu}}
\global\long\def\fb#1{#1^{\left(a\right)}}
\global\long\def\mt#1{#1^{\left(b\right)}}
\global\long\def\phs#1{#1^{\left(p\right)}}
\global\long\def\thc{h}
\global\long\def\state{\mathbf{s}}
\global\long\def\harmonicper{\breve{\matper}}
\global\long\def\kb{k_{B}}
\global\long\def\cb{\bar{c}}
\global\long\def\mb{\bar{\mu}}
\global\long\def\rb{\bar{\rho}}
\global\long\def\wavenumber{k}

\title[Band-gaps in dielectric laminates subjected to incremental shear]{Band-gaps in electrostatically controlled dielectric laminates subjected to incremental shear motions}

\author{Gal Shmuel$^\dagger$ and Gal deBotton$^{\dagger\ddagger\star}$}
\maketitle

\centerline{$^\dagger$Department of Mechanical Engineering}
\centerline{Ben-Gurion University}
\centerline{Beer-Sheva, 84105}
\centerline{Israel}
\vskip 0.2in
\centerline{$^\ddagger$Department of Biomedical Engineering}
\centerline{Ben-Gurion University}
\centerline{Beer-Sheva, 84105}
\centerline{Israel}

\setcounter{footnote}{1}
\footnotetext{\lowercase{E-mail: debotton@bgumail.bgu.ac.il, Tel: int. (972) 8 - 647 7105,
Fax: int. (972) 8 - 647 7106}}

\begin{abstract}
The thickness vibrations of a finitely deformed infinite periodic
laminate made out of two layers of dielectric elastomers is studied.
The laminate is pre-stretched by inducing a bias electric field perpendicular
the the layers. Incremental time-harmonic fields superimposed on the
initial finite deformation are considered next. Utilizing the Bloch-Floquet
theorem along with the transfer matrix method we determine the dispersion
relation which relates the incremental fields frequency and the phase
velocity.

Ranges of frequencies at which waves cannot propagate are identified
whenever the Bloch-parameter is complex. These \emph{band-gap}s depend
on the phases properties, their volume fraction, and most importantly
on the electric bias field. Our analysis reveals how these band-gaps
can be shifted and their width can be modified by changing the bias
electric field. This implies that by controlling the electrostatic
bias field desired frequencies can be filtered out. Representative
examples of laminates with different combinations of commercially
available dielectric elastomers are examined.

\end{abstract}

\section{Introduction}

Band-gaps corresponding to ranges of frequencies at which waves cannot
propagate earned the interest of the scientific community \citep[e.g.,][to name a few]{zieg77ijss,wang&auld85ius,kush&etal94prb,gei&etal04mms,gei&etal09jap}. 
These band-gaps (BGs) or stop-bands can be utilized for various applications such as filtering elastic waves, enhancing performance of ultrasonic transducers, supplying a vibrationless ambient when needed for industrial and biomedical applications, and more. 
Several techniques are available for investigating wave propagation in composites, \citep[e.g., for the
plane-wave method in periodic composites][]{kush&hale93prl,tana&tamu98prb,siga&econ96euphlt} and
the effective medium methods \citep[e.g.,][for composites with random microstructre]{sab&wil88wamo}. As in this work we
are interested in the appearance of BGs in infinite periodic laminates,
we find it convenient to use the transfer matrix method \citep{wang&auld86ieee6},
along with the Bloch-Floquet theorem \citep{kohn&krum72jam}. Specifically,
we consider waves superposed on a pre-deformed state due to bias electric
field acting on an infinite periodic laminate with a repeating sequence of
of dielectric elastomers (DEs) layers. 
The motivation for the choice of DEs as the composite constituents stems from their ability to undergo
large deformations \citep{pelr&etal00scie}, and to change their mechanical and dielectric properties when subjected to electrostatic fields \citep{kofo08jpdp}, thus effecting the way in which electroelastic waves propagate in the matter \citep{shmu&etal11ijnm,gei&etal11iatm}. 
As will be shown in the sequel, \emph{by properly tuning the bias electrostatic field different ranges of
frequencies can be filtered out. }

The structure of this paper is as follows.
Following the works of 
\citet{dorf&ogde05acmc,gdb&etal07mams,dorf&ogde10imajam,bert&gei11jmps,rudy&debo11zamp,ppc&sibo11ijnm} and \citet{tian&etal12jmps}, the background required for describing the static and dynamic responses of elastic dielectric laminates is revisited in section 2. Particularly,
the governing equations for small fields superimposed on large deformations
of electroactive elastomers are summarized. In this section we also
introduce an adequate extension of the transfer matrix method and
the Bloch-Floquet theorem required for treating the propagation of
incremental electroelastic waves superimposed on finitely deformed
DE laminates. Section 3 deals with the response an infinite periodic
laminate with alternating layers, whose behaviors are governed by
the \emph{incompressible dielectric neo-Hookean} model (DH), to a
fixed electric displacement field normal to the layers plane. Subsequently,
the response to small harmonic perturbations superimposed on the finitely
deformed laminate is analyzed. The end result is given in terms of
a dispersion relation relating the incremental fields frequency and
phase velocity to the Bloch-parameter.
Several examples are considered in section 4, where the influence
of the morphology of the laminate, the contrast between the phases
properties, and the bias electric field on the dispersion relation
is examined. Finally, realizable tunable isolators are studied by considering
combinations of commercially available DEs such as VHB 4910, fluorosilicone
730 and ELASTOSIL RT-625.

\section{Theoretical background}

Consider a heterogeneous body occupying a volume region $\Omega_{0}\subset\mathbb{R}^{3}$
made out of $N$ perfectly bonded different homogeneous phases. The
external boundary of the body $\refbnd$ separates it from the surrounding
space $\mathbb{R}^{3}\backslash\Omega_{0}$, assumed to be vacuum.
Each phase occupies a volume region $\Omega_{0}^{\left(r\right)}\left(r=1,2,...,N\right)$
and its boundary is $\partial\Omega_{0}^{\left(r\right)}$. Let $\deformation:\refbody\times\mathcal{I}\rightarrow\mathbb{R}^{3}$
describe a continuous and twice-differentiable mapping of a material
point $\mathbf{X}$ from the reference configuration of the body to
its current configuration $\Omega$ with boundary $\bnd$ by $\sys=\deformation\left(\mathbf{X},t\right)$.
The corresponding velocity and acceleration are denoted, respectively,
by $\velocity=\deformation_{,t}$ and $\accel=\deformation_{,tt},$
and the deformation gradient is $\dg=\Grad\deformation.$
Vectors in the neighborhood of $\mathbf{X}$ are transformed into
vectors in the current configuration via $\mathrm{d}\mathbf{x}=\dg\mathrm{d}\mathbf{X}$.
The volume change of a referential volume element $\mathrm{d}V$ is
given by $\mathrm{d}v=J\mathrm{d}V$, where $\mathrm{d}v$ is the
corresponding volume element in the current configuration, and $J\equiv\det\left(\mathbf{F}\right)>0$
due to material impenetrability. The conservation of mass implies
that $\rho_{L}=J\rho$, where $\rho_{L}$ and $\rho$ are the material
mass densities in the reference and the deformed configurations, respectively.
An area element $\mathrm{d}A$ with the unit normal $\mathbf{N}$
in the reference configuration is transformed to a deformed area $\mathrm{d}a$
with the unit normal $\mathbf{n}$ in the current configuration according
to Nanson's formula $\normall\areal=\frac{1}{J}\dgt\normal\area$.
The right and left Cauchy-Green strain tensors are $\rcg=\dgt\mathbf{F}$
and $\lcg=\dg\dgt$.

The electric field in the current configuration, denoted by $\ef$,
is given in terms of a gradient of a scalar field, termed the electrostatic
potential. In free space an induced electric displacement field $\ed$
is related to the electric field via the vacuum permittivity $\perm$,
such that $\ed=\perm\ef$. In dielectric bodies the connection between
the fields is specified by an adequate constitutive relation. 

In terms of a 'total' stress tensor $\cauchy$ the equations of motion
read 
\begin{equation}
\divg\cauchy=\rho\accel,\label{eq:eom}
\end{equation}
where in order to satisfy the balance of angular momentum symmetry
of $\cauchy$ is required. The terminology 'total' stress is used
to indicate that $\cauchy$ accounts for both mechanical and electrical
contributions. Thus, the traction $\traction$ on a deformed area
element with a unit normal $\normal$ is given by 
\begin{equation}
\cauchy\normal=\traction.\label{eq:cauchy theorem}
\end{equation}
When ideal dielectrics are considered no free body charge is present,
and Gauss's law takes the form
\begin{equation}
\divg\ed=0.\label{eq:gauss law}
\end{equation}
Faraday's law is 
\begin{equation}
\grad\times\ef=\mathbf{0},\label{eq:faraday law}
\end{equation}
when a quasi-electrostatic approximation is considered. A full description
of the system should consider interactions with fields outside the
body, henceforth denoted by a star superscript. Specifically, these
are 
\begin{eqnarray}
\outer{\ed} & = & \perm\outer{\ef},\label{eq:outer d e}\\
\outer{\cauchy} & = & \perm\left[\outer{\ef}\otimes\outer{\ef}-\frac{1}{2}\left(\outer{\ef}\cdot\outer{\ef}\right)\identity\right],\label{eq:maxwell stress}
\end{eqnarray}
where $\identity$ is the identity tensor. In vacuum the governing
Eqs.~\eqref{eq:gauss law}-\eqref{eq:faraday law} for $\outer{\ed}$
and $\outer{\ef}$ are equivalent to Laplace's equation for the electrostatic
potential. As a consequence, $\outer{\cauchy}$ known as the Maxwell
stress, is divergence-free. Across the outer boundary $\bnd$, the
electric jump conditions are 

\begin{eqnarray}
\left(\ed-\outer{\ed}\right)\cdot\normal=-\surfacech, & \left(\ef-\outer{\ef}\right)\times\normal=\mathbf{0},\label{eq:jumps}
\end{eqnarray}
where $\surfacech$ is the surface charge density. In order to formulate
the jump in the stress we postulate a separation of the traction into
a sum of a prescribed mechanical traction $\traction_{m}$, and an
electric traction $\traction_{e}$ induced by the external stress
such that $\traction_{e}=\outer{\cauchy}\normal$. Hence, the stress
boundary condition is

\begin{equation}
\left(\cauchy-\outer{\cauchy}\right)\normal=\traction_{m}.\label{eq:stress bc external}
\end{equation}
The jump conditions across a charge-free internal boundary between
two adjacent phases $m$ and $f$ are
\begin{eqnarray}
\jump{\cauchy}\normal=\mathbf{0}, & \jump{\ed}\cdot\normal=0, & \jump{\ef}\times\normal=\mathbf{0},\label{eq:jumps-internal}
\end{eqnarray}
where $\jump{\bullet}\equiv\left(\bullet\right)^{\left(b\right)}-\left(\bullet\right)^{\left(a\right)}$
denotes the jump of fields between the two phases. 

A Lagrangian formulation of the governing equations and jump conditions
is feasible by a \emph{pull-back} of the Eulerian fields 
\begin{eqnarray}
\piola=J\cauchy\idgt, & \edl=J\idg\ed, & \efl=\dgt\ef,\label{eq:pull back p d e}
\end{eqnarray}
for the 'total' first Piola-Kirchhoff stress, Lagrangian electric
displacement and electric field, respectively. Thus, the governing
Eqs.~\eqref{eq:eom}, \eqref{eq:gauss law} and \eqref{eq:faraday law}
transform, respectively, to 
\begin{eqnarray}
\Div\cdot\piola=\rho_{L}\accel, & \Div\cdot\edl=0, & \Curl\times\efl=\mathbf{0}.\label{eq:l eom d e}
\end{eqnarray}
The corresponding referential jump conditions across the external
boundary $\refbnd$ are 
\begin{eqnarray}
\left(\piola-\outer{\piola}\right)\normall=\traction_{M}, & \left(\edl-\outer{\edl}\right)\cdot\normall=-w_{E}, & \left(\efl-\outer{\efl}\right)\times\normall=\mathbf{0},\label{eq:lagrng jumps}
\end{eqnarray}
where $\traction_{M}\areal=\traction_{m}\area$ and $w_{E}\areal=\surfacech\area$.
Across the referential interfaces Eqs.~\eqref{eq:jumps-internal}
become 
\begin{eqnarray}
\jump{\piola}\normall=0, & \quad\jump{\edl}\cdot\normall=0, & \quad\jump{\efl}\times\normall=\mathbf{0}.\label{eq:lagrng internal jumps}
\end{eqnarray}

Following \citet{dorf&ogde05acmc}, the constitutive relation is expressed in terms of an \emph{augmented energy density function }(AEDF) $\aef$ with the independent variables $\dg$ and $\edl$, such that the total first Piola-Kirchhoff stress and the Lagrangian electric field are derived via
\begin{equation}
\piola=\derivative{\aef}{\dg},\quad\efl=\derivative{\aef}{\edl}.\label{eq:constitutive law}
\end{equation}
The first of Eq. (\ref{eq:constitutive law}) should be modified when
considering incompressible materials, as the kinematic constraint
yields an additional workless part of stress. The latter is accounted
by introducing a Lagrange multiplier $p_{0}$ which can determined
only from the equilibrium equations together with the boundary conditions.
Thus, the total first Piola-Kirchhoff is 
\begin{equation}
\piola=\derivative{\aef}{\dg}-p_{0}\idgt.\label{eq:stress for inc}
\end{equation}

Based on the framework developed in \citet{dorf&ogde10imajam}, we  superimpose
an infinitesimal time-dependent elastic and electric displacements
$\inc{\sys}=\inc{\deformation}\left(\refsys,t\right)$ and $\inc{\edl}\left(\refsys,t\right)$,
on the pre-deformed configuration. 
Herein, and in the sequel, we use a superposed dot to denote incremental
quantities. Let the Eulerian quantities $\piolaincpush,\edlincpush,\ \mathrm{and}\ \efincpush$
denote the \emph{push-forwards }of increments in the first Piola-Kirchhoff
stress, the Lagrangian electric displacement and electric fields,
respectively, namely

\begin{equation}
\piolaincpush=\frac{1}{J}\inc{\piola}\dg^{T},\quad\edlincpush=\frac{1}{J}\dg\inc{\edl},\quad\efincpush=\dg^{-T}\inc{\efl}.
\end{equation}
In terms of these variables, the incremental governing equations read
\begin{eqnarray}
\divg\piolaincpush=\rho\inc{\sys}_{,tt}, & \quad\divg\edlincpush=0, & \quad\grad\times\efincpush=0.\label{eq:inc eom gauss farady current conf}
\end{eqnarray}
For an incompressible material the incremental constraint is 
\begin{equation}
\divg\inc{\sys}\equiv\tr{\disgrad}=0,\label{eq:inc compressibility}
\end{equation}
where $\disgrad=\grad\inc{\sys}$ is the displacement gradient. Linearization
of the incompressible material constitutive equations in the increments
yields 
\begin{eqnarray}
\piolaincpush & = & \elaspush\disgrad+p_{0}\disgrad^{\mathrm{T}}-\inc p_{0}\identity+\elaselecpush\edlincpush,\label{eq:constitutive eq for the push of piola}\\
\efincpush & = & \elaselecpush^{\mathrm{T}}\disgrad+\elecpush\edlincpush,\label{eq:constitutive eq for the push of inc E}
\end{eqnarray}
where $\left(\elaselecpush^{\mathrm{T}}\disgrad\right)_{k}=\elaselecpushcomp{ijk}h_{ij}$.
The quantities $\elecpush,\elaselecpush$ and $\elaspush$ are the
push-forwards of the referential reciprocal dielectric tensor, electroelastic
coupling tensor, and elasticity tensor, respectively. In components
form 
\begin{eqnarray}
\elecpushcomp{ij}=J\idgcomp{\alpha i}\elecpushcomp{0\alpha\beta}\idgcomp{\beta j}, & \elaselecpushcomp{ijk}=\dgcomp{j\alpha}\elaselecpushcomp{0i\alpha\beta}\idgcomp{\beta k}, & \elaspushcomp{ijkl}=\frac{1}{J}\dgcomp{j\alpha}\elaspushcomp{0i\alpha k\beta}\dgcomp{l\beta},\label{eq:A B C push def}
\end{eqnarray}
where\begin{equation}
\label{A0B0C0comp}
\elecpushcomp{0\alpha\beta}=\derivative{^{2}\aef}{\edlcomp{\alpha}\partial\edlcomp{\beta}},\quad\elaselecpushcomp{0i\alpha\beta}=\derivative{^{2}\aef}{\dgcomp{i\alpha}\partial\edlcomp{\beta}},\quad\elaspushcomp{0i\alpha k\beta}=\derivative{^{2}\aef}{\dgcomp{i\alpha}\partial\dgcomp{k\beta}}.
\end{equation}The incremental outer fields are given by 
\begin{eqnarray}
\inc{\outer{\ed}} & = & \perm\inc{\outer{\ef}},\label{eq:in electro outer fields}\\
\inc{\outer{\cauchy}} & = & \perm\left[\inc{\outer{\ef}}\otimes\outer{\ef}+\outer{\ef}\otimes\inc{\outer{\ef}}-\left(\outer{\ef}\cdot\inc{\outer{\ef}}\right)\identity\right].\label{eq:inc maxwell stress}
\end{eqnarray}
Here again $\inc{\outer{\ed}}$ and $\inc{\outer{\ef}}$ are to satisfy
Eqs.~\eqref{eq:gauss law}-\eqref{eq:faraday law}, where subsequently
$\divg\inc{\outer{\cauchy}}=\mathbf{0}$.

The push-forward of the increments of Eqs.~\eqref{eq:lagrng jumps}
gives the following incremental jump conditions across the external
boundary in the current configuration 
\begin{eqnarray}
\left[\piolaincpush-\inc{\outer{\cauchy}}+\outer{\cauchy}\trans{\disgrad}-\left(\divg\inc{\sys}\right)\outer{\cauchy}\right]\normal & = & \check{\traction}_{m},\label{eq:in stress jump}\\
\left[\edlincpush-\inc{\outer{\ed}}-\left(\divg\inc{\sys}\right)\outer{\ed}+\disgrad\outer{\ed}\right]\cdot\normal & = & -\check{w}_{e},\label{eq:in charge  jump}\\
\left[\mathrm{\efincpush}-\inc{\outer{\ef}}-\trans{\disgrad}\outer{\ef}\right]\times\normal & = & \mathbf{0},\label{eq:inc elec jump}
\end{eqnarray}
where $\check{\traction}_{m}\mathrm{d}a=\inc{\traction}_{M}\mathrm{d}A$
and $\check{w}_{e}\mathrm{d}a=\inc w_{E}\mathrm{d}A$. Similarly,
the push forward of the increments of Eqs.~\eqref{eq:lagrng internal jumps}
results in the following jumps across the internal boundaries 
\begin{eqnarray}
\jump{\piolaincpush}\normal=\mathbf{0}, & \jump{\edlincpush}\cdot\normal=0 & \jump{\mathrm{\efincpush}}\times\normal=\mathbf{0}.\label{eq:internal inc jump current}
\end{eqnarray}

Several techniques are available in the literature for tackling the
problem of wave propagation in periodic composites.  In layered structures
it is convenient to utilize the transfer-matrix method in conjunction
with the Bloch-Floquet theorem \citep{adle90uffcieee}. Herein we formulate
an adequate adjustment for treating the propagation of incremental
electroelastic waves superimposed on a finitely deformed multiphase
DE laminates. 

\begin{figure}[t]
\includegraphics[scale=0.7,angle=0]{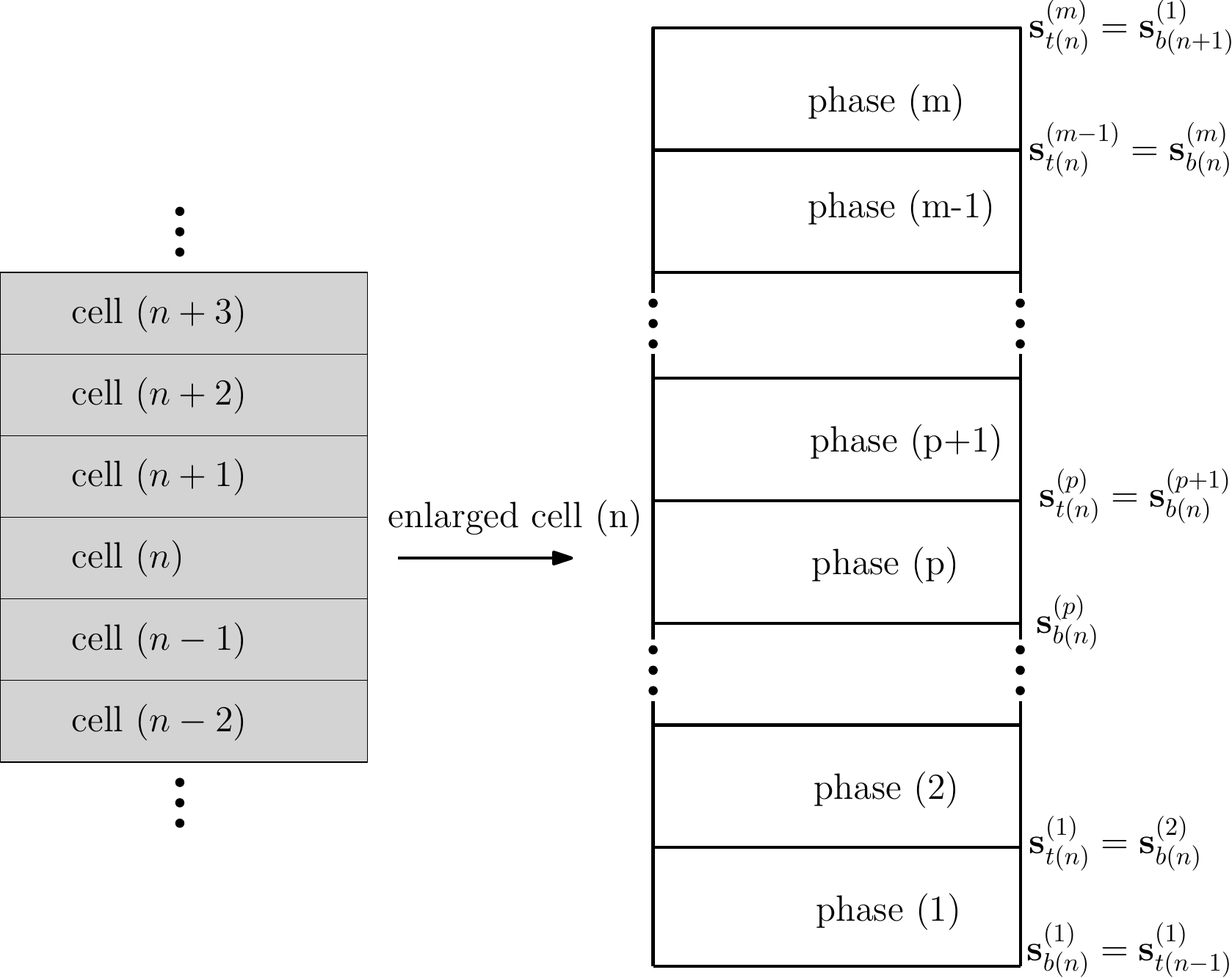}
\caption{An illustration of an infinite laminate composed  of periodic $m$-layers unit-cells.}
\label{tmm}
\end{figure}

Consider a finitely deformed infinite DE-laminate made out of $m$-phase
periodic unit-cells subjected to incremental perturbations harmonic
in time and normal to the layers plane. 
A schematic illustration of the laminate is shown in Fig.~\ref{tmm}.
Let $\phs{\state_{n}}$, $p=1,..,m$, denote an incremental state
vector in the $p$-phase of the $n^{th}$ cell consisting quantities
which are continuous across the interface between the two adjacent
layers. In view of the linearity of the incremental problem it can
be shown that the state vector at the top of the $p$-phase $\phs{\state_{t(n)}}$
and the state vector at the bottom of the $p$-phase $\phs{\state_{b(n)}}$
(see Fig.~\ref{tmm}) are related via a unique non-singular transfer
matrix $\phs{\mathbf{T}}$ such that 
\begin{equation}
\phs{\state_{t(n)}}=\phs{\mathbf{T}}\phs{\state_{b(n)}}.\label{eq:transfer p}
\end{equation}
 In terms of the state vectors, the continuity conditions across the
interface are 
\begin{equation}
\p{\state_{t(n)}}=\state_{b(n)}^{\left(p+1\right)}.\label{eq:continuity}
\end{equation}
Utilizing Eqs.~\eqref{eq:transfer p}-\eqref{eq:continuity} recursively
the relation between $\state_{t}^{\left(m\right)}$ and $\state_{b}^{\left(1\right)}$
is 
\begin{equation}
\mt{\state_{t(n)}}=\mt{\mathbf{T}}\mt{\state_{b(n)}}=\mt{\mathbf{T}}\state_{t(n)}^{\left(m-1\right)}=\mt{\mathbf{T}}\mathbf{T}^{\left(m-1\right)}\state_{b(n)}^{\left(m-1\right)}=...=\mathbf{T}\state_{b}^{\left(1\right)},\label{eq:total transfer}
\end{equation}
where $\mathbf{T}\equiv\prod_{p=1}^{m}\mathbf{T}^{\left(p\right)}$
is the total transfer matrix.

At one hand the continuity conditions across the interface between
two successive cells $n$ and $n+1$ are 
\begin{equation}
\state_{b\left(n+1\right)}^{\left(1\right)}=\state_{tn}^{\left(m\right)}.\label{eq:continuity between cells}
\end{equation}
At the other, the Bloch-Floquet's theorem states  that the state-vectors
of the same phase in adjacent cells are identical up to a phase shift
in the form 
\begin{equation}
\phs{\state_{b\left(n+1\right)}}=e^{-ik_{B}h}\phs{\state_{b\left(n\right)}},\label{eq:bloch condition}
\end{equation}
where $0\leq k_{B}<\pi/h$ is the Bloch-Floquet wavenumber in first
irreducible Brillouin zone, representing the smallest region where
wave propagation is unique \citep{kitt05book}. Substitution of
Eq.~\eqref{eq:total transfer} in \eqref{eq:continuity between cells},
followed by employing the latter with Eq.~\eqref{eq:bloch condition},
yields the eigenvalue problem 
\begin{equation}
\det\left|\mathrm{T}-e^{-ik_{B}h}\mathrm{I}\right|=0,\label{eq:dispersion relation}
\end{equation}
where $\mathrm{I}$ is an identity matrix with the dimension of the
total transfer matrix. The solution of Eq.~\eqref{eq:dispersion relation}
provides the dispersion relation describing the manner by which waves
propagate in the laminate.

\section{Thickness vibrations of an infinite periodic two-phase dielectric
laminate}

Henceforth we restrict our attention to isotropic phases.
Consequently, the phases AEDFs can be written in terms of the six invariants \citep{dorf&ogde05acmc} 
\begin{eqnarray}
I_{1}=\tr{\rcg}=\rcg:\mathbf{I}, & I_{2}=\frac{1}{2}\left(I_{1}^{2}-\rcg:\rcg\right), & I_{3}=\det\left(\rcg\right)=J^{2},\label{eq:c invariants}\\
\ii{4e}=\edl\cdot\edl, & \ii{5e}=\edl\cdot\rcg\edl, & \ii{6e}=\edl\cdot\rcg^{2}\edl.\label{eq:d invariants}
\end{eqnarray}
Specifically, we consider phases which are characterized by \emph{incompressible
dielectric neo-Hookean} (DH) model 
\begin{equation}
\phase{\dnh}\left(\ii 1,\ii{5e}\right)=\frac{\phase{\mu}}{2}\left(\ii 1-3\right)+\frac{1}{2\phase{\matper}}\ii{5e},\label{eq:dielecric neo hookean}
\end{equation}
where $\phase{\mu}$ and $\phase{\matper}=\perm\phase{\relper}$ denote
the phases shear moduli and dielectric constants, respectively, and
$\phase{\relper}$ being the phases relative dielectric constant.
The corresponding total stress is 
\begin{equation}
\phase{\cauchy}=\phase{\mu}\lcg+\frac{1}{\phase{\matper}}\ed\otimes\ed-\phase{p_{0}}\identity,\label{eq:d nh stress}
\end{equation}
and the current displacement and electric field are related via the
isotropic linear relation 
\begin{equation}
\ed=\phase{\matper}\ef.\label{eq:de of DH}
\end{equation}
Calculation of the associated constitutive tensors $\phase{\elecpush},\phase{\elaselecpush},\ \mathrm{and\ }\phase{\elaspush}$
yields, in components form,

\begin{equation}
\phase{\elecpushcomp{ij}}=\frac{1}{\p{\matper}}\cronck{ij},\quad\phase{\elaselecpushcomp{ijk}}=\frac{1}{\phase{\matper}}\left(\cronck{ik}\edcomp j+\frac{\edcomp i}{\phase{\matper}}\cronck{jk}\right),\quad\phase{\elaspushcomp{ijkl}}=\phase{\mu}\cronck{ik}\lcgcomp{jl}+\frac{1}{\phase{\matper}}\cronck{ik}\edcomp j\edcomp l,\label{eq:dnh a b  c comp}
\end{equation}
where $\cronck{ij}$ are the components of the Kronecker delta.

\begin{figure}[t]
\includegraphics[scale=0.7,angle=0]{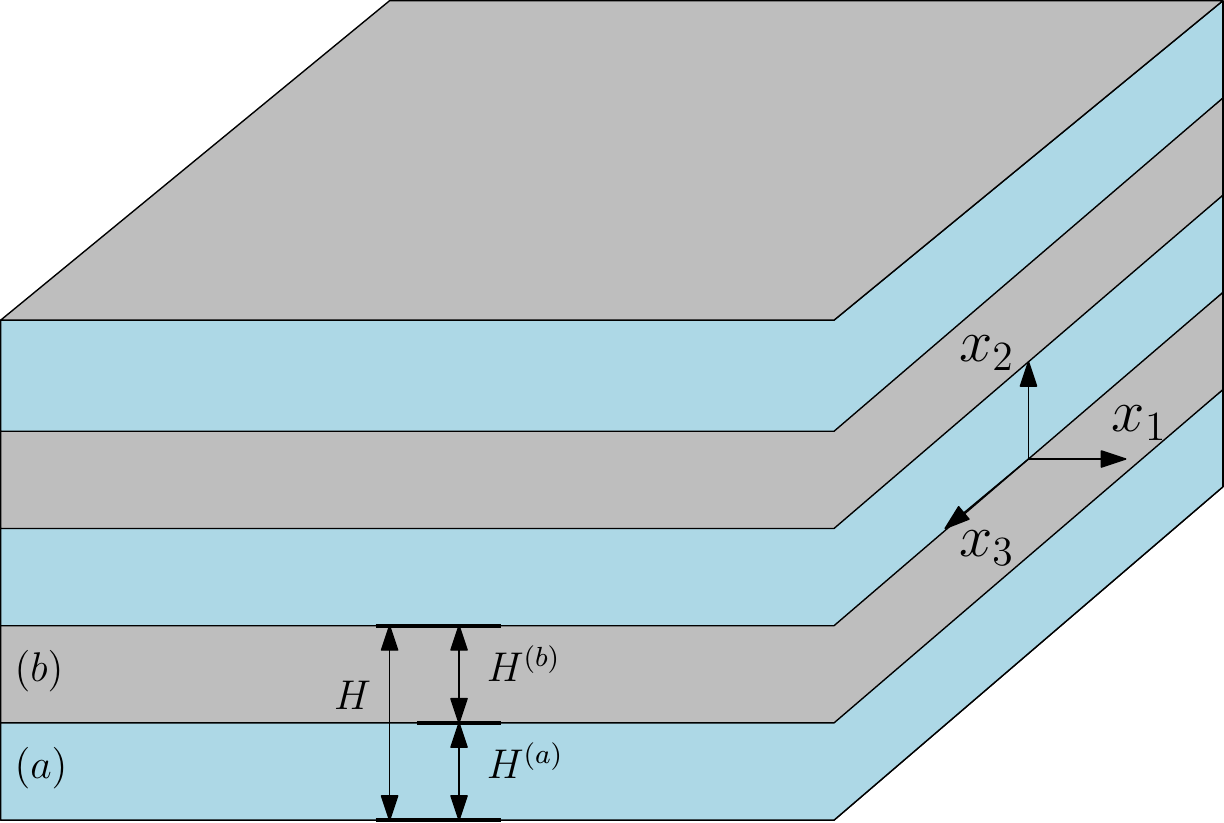}
\caption{An illustration of an infinite periodic laminate, composed  of alternating $a$ and $b$ phases, with initial thicknesses  $H^{(a)}$ and $H^{(b)}$, respectively. The   electric bias field is  along the $x_{2}$ direction.}
\label{laminate}
\end{figure}

Consider a Cartesian coordinate system with unit vectors $\mathbf{i_{1},i_{2}}$
and $\mathbf{i_{3}}$ along $x_{1},x_{2}$ and $x_{3}$, respectively.
Let a DE laminate be composed of periodic two-phase unit-cells with
thickness $H$ along the the $x_{2}$ direction, infinite along the
$x_{1}$ and $x_{3}$ directions as illustrated in Fig.~\ref{laminate}.
For convenience we denote the phases $a$ and $b$. The sum of the
phases thicknesses equals the unit-cell thickness $H=\fb H+\mt H$,
such that $\fb v=\fb H/H$ and $\mt v=\mt H/H$ are the volume fractions
of phase $a$ and phase $b$, respectively. We assume that the composite
is allowed to expand freely in the $x_{1}$ and $x_{3}$ directions.
Of interest is the composite response when subjected to an overall
electric displacement field $\bar{\ed}=d_{2}\mathbf{i_{2}}$, where
$d_{2}\equiv\fb v\fb{d_{2}}+\mt v\mt{d_{2}}$. 

In each phase we assume displacement fields compatible with the homogeneous
uni-modular (on account of incompressibility) diagonal deformation
gradients 
\begin{equation}
\phs{\left[\mathrm{F}\right]}=\mathrm{diag}\left[\phs{\lambda_{1}},\phs{\lambda_{2}},\left(\phs{\lambda_{1}}\phs{\lambda_{2}}\right)^{-1}\right],\label{eq:defromation gradient p f,m}
\end{equation}
where henceforth $p=a,b$. In virtue of the perfect bonding between
the phases, the stretch ratios in the $x_{1}$ and $x_{3}$ directions
in the phases are the same, i.e.,
\begin{eqnarray}
\mt{\lambda_{1}}=\fb{\lambda_{1}}, & \left(\mt{\lambda_{1}}\mt{\lambda_{2}}\right)^{-1}=\left(\fb{\lambda_{1}}\fb{\lambda_{2}}\right)^{-1}.\label{eq:lflm}
\end{eqnarray}
Eqs.~\eqref{eq:lflm} imply that $\mt{\lambda_{2}}=\fb{\lambda_{2}}$,
and henceforth we define $\lambda\equiv\fb{\lambda_{1}}$. 
From the symmetry of the problem in the $x_{1}-x_{3}$ it follows that $\p{\lambda_{2}}=\lambda^{-2}$. 
The corresponding stresses in each phase are 
\begin{eqnarray}
\p{\cauchycomp{11}}=\p{\cauchycomp{33}}=\p{\mu}\lambda^{2}-\p{p_{0}}, 
& 
\quad\p{\cauchycomp{22}}=\frac{\p{\mu}}{\lambda^{4}}+\frac{1}{\p{\epsilon}}(\edcomp 2^{(p)})^{2}-\p{p_{0}}.
\label{eq:stress components bias}
\end{eqnarray}

We assume mechanical traction-free boundary conditions at infinity.
Utilizing the jump conditions in the first of (\ref{eq:jumps}) and
\eqref{eq:stress bc external}, along with the traction-free boundary
conditions we have that 
\begin{eqnarray}
\mt{\edcomp 2}=\fb{\edcomp 2}=d_{2}, & \mt{\cauchycomp{22}}=\fb{\cauchycomp{22}}=0.\label{eq:jump in finite laminate}
\end{eqnarray}
From the latter together with the second of Eq. (\ref{eq:stress components bias})
the expressions for the pressure in the phases are 
\begin{equation}
\phs{p_{0}}=\frac{\phs{\mu}}{\lambda^{4}}+\frac{d_{2}^{2}}{\phs{\matper}}.\label{eq:p l}
\end{equation}
Substitution of Eq.~\eqref{eq:p l} into the first of Eq. (\ref{eq:stress components bias})
leads to a relation between the applied Lagrangian electric displacement
field $\edl=\lambda^{2}\ed$ and the resultant stretch ratio, namely
\begin{equation}
\lambda=\left(1+D_{2}^{2}\frac{\harmonicper}{\bar{\mu}}\right)^{1/6},\label{eq:l d}
\end{equation}
where $\harmonicper=\fb v/\fb{\matper}+\mt v/\mt{\matper}$ and $\mb=\fb{\mu}\fb v+\mt{\mu}\mt v$.
The latter appropriately reduces to the corresponding relation for
a homogeneous specimen. In terms of the dimensionless quantities $\alpha=\fb{\mu}/\mt{\mu}$,
$\beta=\fb{\matper}/\mt{\matper}$, and $\dh=\edlcomp 2/\sqrt{\mt{\mu}\mt{\matper}}$,
denoting the shear contrast, permittivity contrast, and normalized
Lagrangian electric displacement field, respectively, the stretch
ratio is 
\begin{eqnarray}
\lambda & = & \left(1+\hat{D}^{2}\frac{\mt v+\fb v\beta^{-1}}{\mt v+\fb v\alpha}\right)^{1/6}.\label{eq:l d p d}
\end{eqnarray}
 
For the above choice of constitutive relation, we also have that Eq.~\eqref{eq:dnh a b  c comp}
reduces to
\begin{eqnarray}
\p{\elecpushcomp{11}} & = & \p{\elecpushcomp{22}}=\p{\elecpushcomp{33}}=\frac{1}{\p{\matper}},\\
\p{\elaselecpushcomp{121}} & = & \p{\elaselecpushcomp{211}}=\p{\elaselecpushcomp{323}}=\p{\elaselecpushcomp{233}}=\frac{1}{2}\p{\elaselecpushcomp{222}}=\frac{1}{\p{\matper}}\p{\edcomp 2},\\
\p{\elaspushcomp{1111}} & = & \p{\elaspushcomp{2121}}=\p{\elaspushcomp{2323}}=\p{\elaspushcomp{1313}}=\p{\elaspushcomp{3333}}=\p{\mu}\lambda^{2},\\
\p{\elaspushcomp{1212}} & = & \p{\elaspushcomp{2222}}=\p{\elaspushcomp{3232}}=\p{\mu}/\lambda^{4}+\frac{1}{\p{\epsilon}}\edcomp 2^{\left(p\right)2}.
\end{eqnarray}

The response of the laminate to a harmonic excitation superimposed
on the aforementioned finite deformation is addressed next. Let $\phs{\inc x}_{i}$
and $\phs{\iee}$ denote, respectively, the components of the incremental
displacement and the incremental electric potential in the phases, such that $\phs{\mathrm{\efincpush}}=-\grad\phs{\iee}$.
The fields sought are to satisfy the incremental equations of motion
along with Gauss equation (\ref{eq:inc eom gauss farady current conf})
and the incompressibility constraint \eqref{eq:inc compressibility}.
Following \citet{tier63jasa}, we consider fields that are functions of $x_{2}$ and time alone, independent of $x_{1}$ and $x_{3}$, compatible with the simple-thickness mode \citep{mind55book}.
Consequently, in each phase the governing equations (\ref{eq:inc eom gauss farady current conf})
and \eqref{eq:inc compressibility} simplify to
\begin{subequations}
\begin{align}
\label{eomx}
\p{\effectivemu}\p{\disgradcomp{12,2}}&=\p{\rho}\p{\inc x_{1,tt}},\\
\label{eomy}
\left(\p{\effectivemu}-\frac{3}{\p{\matper}}\edcomp 2^{2}+\p{p_{0}}\right)\p{\disgradcomp{22,2}}-\p{\inc p_{0,2}}-2\edcomp 2\p{\iee_{,22}}&=\p{\rho}\p{\inc x_{2,tt}},\\
\label{eomz}
\p{\effectivemu}\p{\disgradcomp{32,2}}&=\p{\rho}\p{\inc x_{3,tt}},
\end{align}
\end{subequations}
and
\begin{subequations}
\begin{align}
\label{gaussinc}
\p{\edlincpushcomp{2,2}}&=0,\\
\label{inccomp}
\p{\disgradcomp{22}}&=0,
\end{align}
\end{subequations}
respectively, where $\p{\effectivemu}=\p{\mu}/\lambda^{4}$, $\p{\edlincpushcomp 2}=-\p{\matper}\left(\edcomp 2\p{\iee_{,2}}+\frac{2}{\p{\matper}}\edcomp 2\p{\disgradcomp{22}}\right)$.
Substitution of Eq.~\eqref{inccomp} into Eq.~\eqref{gaussinc} reveals
that $\p{\iee_{,22}}=0$, hence the incremental electric potential
is linear in $x_{2}$, say $\p{\iee}=\p{L_{1}}+\p{x_{2}}\p{L_{2}}$,
where $\p{L_{1}}$ and $\p{L_{2}}$ are integration constants.
More significantly, it implies that the incremental electrical and
mechanical fields are not coupled, as Eq.~\eqref{eomy} remains a
function of $\p{\disgrad}$ alone, independent of $\p{\varphi}$,
together with Eq.~\eqref{gaussinc} which remains a function of $\p{\varphi}$
alone, independent of $\p{\disgrad}$. 
Eq.~\eqref{inccomp} states that along the thickness the incremental displacement is spatially constant. Together with Eqs.~\eqref{eomx} and \eqref{eomz} the solution for the incremental problem is
\begin{eqnarray}
\phs{\dot{x}}_{1} & = & \left(\phs U_{1}\sin\phs{\wavenumber}x_{2}+\phs U_{2}\cos\phs{\wavenumber}x_{2}\right)e^{-i\afreq t},\nonumber \\
\phs{\dot{x}}_{2} & = & \phs U_{3}e^{-i\afreq t},\label{eq:xd1 xd3}\\
\phs{\dot{x}}_{3} & = & \left(\phs U_{4}\sin\phs{\wavenumber}x_{2}+\phs U_{5}\cos\phs{\wavenumber}x_{2}\right)e^{-i\afreq t},\nonumber 
\end{eqnarray}
where $\phs{\wavenumber}=\afreq/\phs{\bar{c}}$, $\phs{\bar{c}}=\phs{\effectivemu}/\phs{\rho}$,
$\afreq$ is the angular excitation frequency, and ${\phs U_{r}} \left(r=1, 2, \ldots, 5\right)$
are integration constants. Without loss of generality the constants
$\phs U_{3}$ can be set to be zero. 
The motion described by the solution Eq.~\eqref{eq:xd1 xd3}
corresponds to displacement gradients which retain only shear components.
For completeness, the solution for the incremental pressure emerges
from Eq.~\eqref{eomy}, which reduces to $-\p{\inc p_{0,2}}=\p{\rho}\p{\inc x_{2,tt}},$
and implies that $\p{\inc p_{0}}=\p Pe^{-i\afreq t}$, where $\p P$
are integration constants. 
The incremental state vector of the $n^{th}$ cell takes the form 
\begin{equation}
\phs{\mathrm{s}_{(n)}}=\left\{ \phs{\dot{x}}_{n1},\phs{\dot{x}}_{n3},\phs{\piolaincpushcomp{n12}},\phs{\piolaincpushcomp{n22}},\phs{\piolaincpushcomp{n32}},\phs{\iee_{n}},\phs{\edlincpushcomp{n2}}\right\} {}^{\mathrm{T}}.\label{eq:state n p}
\end{equation}
The components of transfer matrix $\phs{\mathbf{T}_{(n)}}$ relating the top $\phs{\state_{t\left(n\right)}}$
and bottom $\phs{\state_{b\left(n\right)}}$ state vectors are given
in the Appendix.
Note that the continuity conditions for $\phs{\dot{x}}_{n2}$
is identically satisfied when we set $\phs U_{3}$ to zero. 
The corresponding solution of the eigenvalue problem Eq.~\eqref{eq:dispersion relation}
resulting from the Bloch-Floquet theorem can be expressed in terms
of the trigonometric expression
\begin{equation}
\kb=\frac{1}{h}\arccos\eta,\label{eq:sol of kb}
\end{equation}
where 
\begin{equation}
\eta=\frac{1}{2h}\left[\cos\mt{\wavenumber}\mt h\cos\fb{\wavenumber}\fb h-\frac{1}{2}\left(\mt{\wavenumber}/\fb{\wavenumber}+\fb{\wavenumber}/\mt{\wavenumber}\right)\sin\mt{\wavenumber}\mt h\sin\fb{\wavenumber}\fb h\right],
\label{eq:dispersion sol of eta}
\end{equation}
with $\p h=\p H/\lambda^{2}$ and $h=\fb h+\mt h$. Eq.~\eqref{eq:sol of kb}
is an extension of the solution given in \citet{wang&etal86p1ius} to the
class of finitely deformed infinite periodic DE laminates. The dependency
on the bias field becomes evident when $\eta$ is rephrased in terms
of the electric bias field, the referential geometry and the frequency,
namely 
\begin{eqnarray}
\eta & = &\frac{1}{2H} \left(1+D_{2}^{2}\frac{\harmonicper}{\bar{\mu}}\right)^{1/3}
\bigg[\cos\frac{\afreq\rho^{\left(b\right)}H^{\left(b\right)}}{\mu^{\left(b\right)}\left(1+D_{2}^{2}\frac{\harmonicper}{\bar{\mu}}\right)^{1/3}}\cos\frac{\afreq\rho^{\left(a\right)}H^{\left(a\right)}}{\mu^{\left(a\right)}\left(1+D_{2}^{2}\frac{\harmonicper}{\bar{\mu}}\right)^{1/3}}\nonumber \\
 &  & -\frac{1}{2}\left(\frac{\alpha\rho^{\left(b\right)}}{\rho^{\left(a\right)}}+\frac{\rho^{\left(a\right)}}{\alpha\rho^{\left(b\right)}}\right)\sin\frac{\afreq\rho^{\left(b\right)}H^{\left(b\right)}}{\mu^{\left(b\right)}\left(1+D_{2}^{2}\frac{\harmonicper}{\bar{\mu}}\right)^{1/3}}\sin\frac{\afreq\rho^{\left(a\right)}H^{\left(a\right)}}{\mu^{\left(a\right)}\left(1+D_{2}^{2}\frac{\harmonicper}{\bar{\mu}}\right)^{1/3}}
 \bigg].
 \label{eq:eta of d}
\end{eqnarray}
In virtue of the uncoupled nature of the incremental problem, the incremental
electric quantities indeed do not appear in Eq.~\eqref{eq:eta of d}. 
We conclude this section noting that waves of certain frequencies cannot propagate when the pertained solution of $k_{B}$ is complex.
These cases will be examined in the following section.

\section{Numerical examples}

\begin{figure}[t]
\includegraphics[scale=0.60]{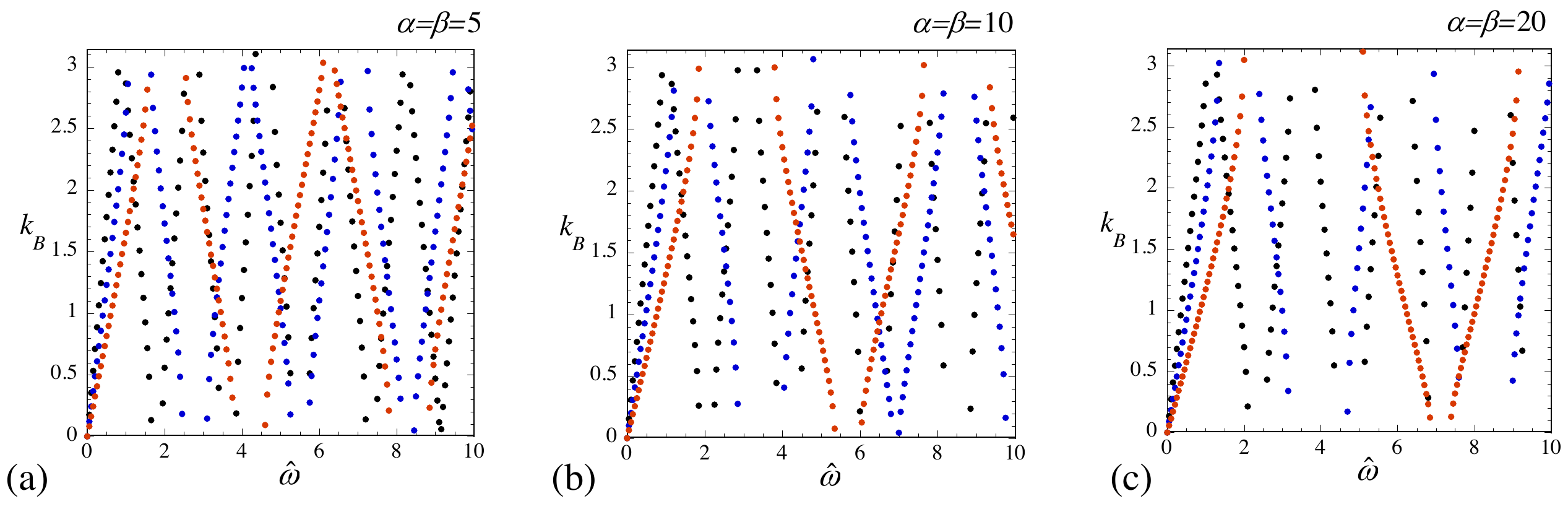} 
\caption{Variations of the Bloch wavenumber $\kb$ as functions of the normalized frequency
$\afreqh=\afreq H/\cb$, for (a) $\alpha=\beta=5$, (b) $\alpha=\beta=10$ and (c) $\alpha=\beta=20$.
The red, blue and black curves correspond to $\mt v=0.2,\ 0.5$ and
0.8, respectively.}
\label{dispersion of alpha}
\end{figure}

\begin{figure}[t]
\includegraphics[angle=0,scale=0.62]{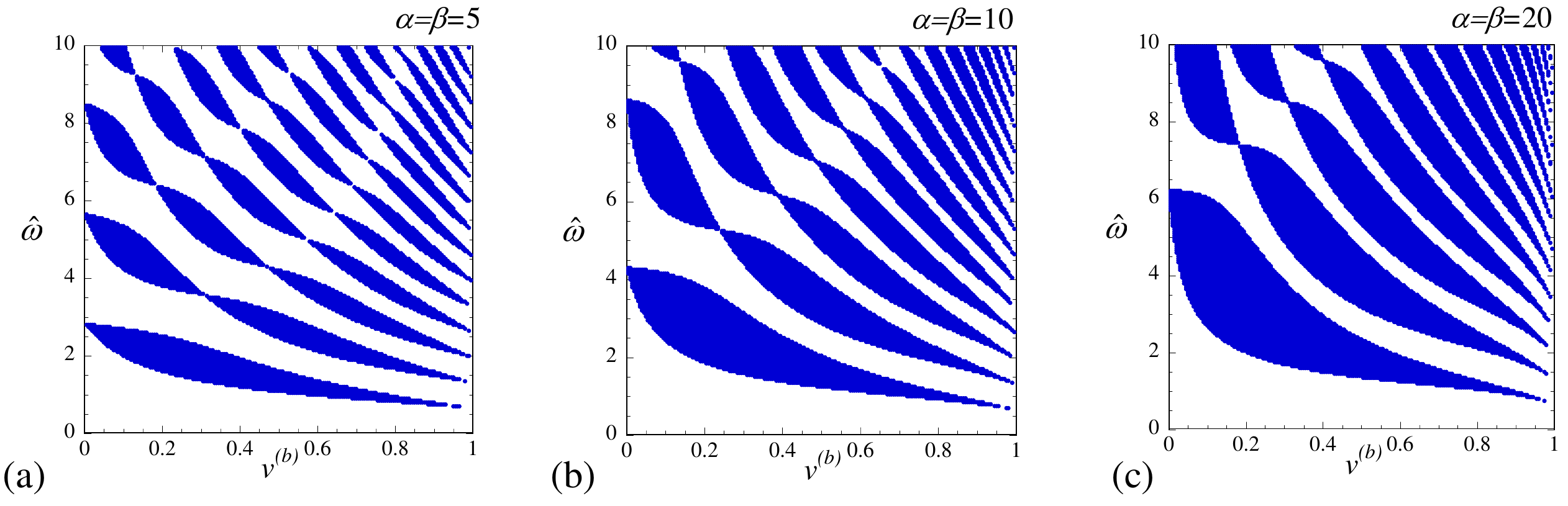} 
\caption{Prohibited frequencies as functions of $\mt v$ at $\dh=3$, for (a)
$\alpha=\beta=5$ and (b) $\alpha=\beta=10$ and (c) $\alpha=\beta=20$.}
\label{bg of vm}
\end{figure}

With the help of a few examples we explore the dispersion relation
characterizing the dynamic response of the composite. We examine its
dependency on the phases volume fractions and the contrast between
the phases shear and dielectric moduli. In particular we are interested
in the influence of the electric bias field on the dynamic response
of the composite. 
To this end we make the following choice for the properties of phase $b$ 
\begin{eqnarray}
\mt{\rho}=1000\left[\frac{kg}{m^{3}}\right], & \mt{\mu}=200\left[KPa\right], & \mt{\relper}=2,
\end{eqnarray}
and determine dispersion diagrams for various choices of phase
$a$ properties and volume fractions at different values of the dimensionless
bias field $\dh$. 
In all the forthcoming examples $\fb{\rho}=\mt{\rho}$
is assumed. 

Fig.~\ref{dispersion of alpha} shows variations of the Bloch wavenumber $\kb$
as functions of the normalized frequency $\afreqh=\afreq H/\cb$ where
$\cb=\sqrt{\mb/\rb}$, $\rb=\fb{\rho}\fb v+\mt{\rho}\mt v$, for a
representative value of the bias field $\dh=3$. 
Figs.~\ref{dispersion of alpha}a, b and c correspond to $\alpha=\beta=5,10$ and 20, respectively. 
The red, blue and black curves correspond to $\mt v=0.2,0.5$ and 0.8,
respectively. 
The period of the dispersion curves
becomes smaller with an increase of the softer phase volume fraction.
An inverse effect is revealed when the contrast is enhanced, that
is longer periods for higher values of $\alpha$ and $\beta$. 

\begin{figure}[t]
\includegraphics[angle=0,scale=0.45]{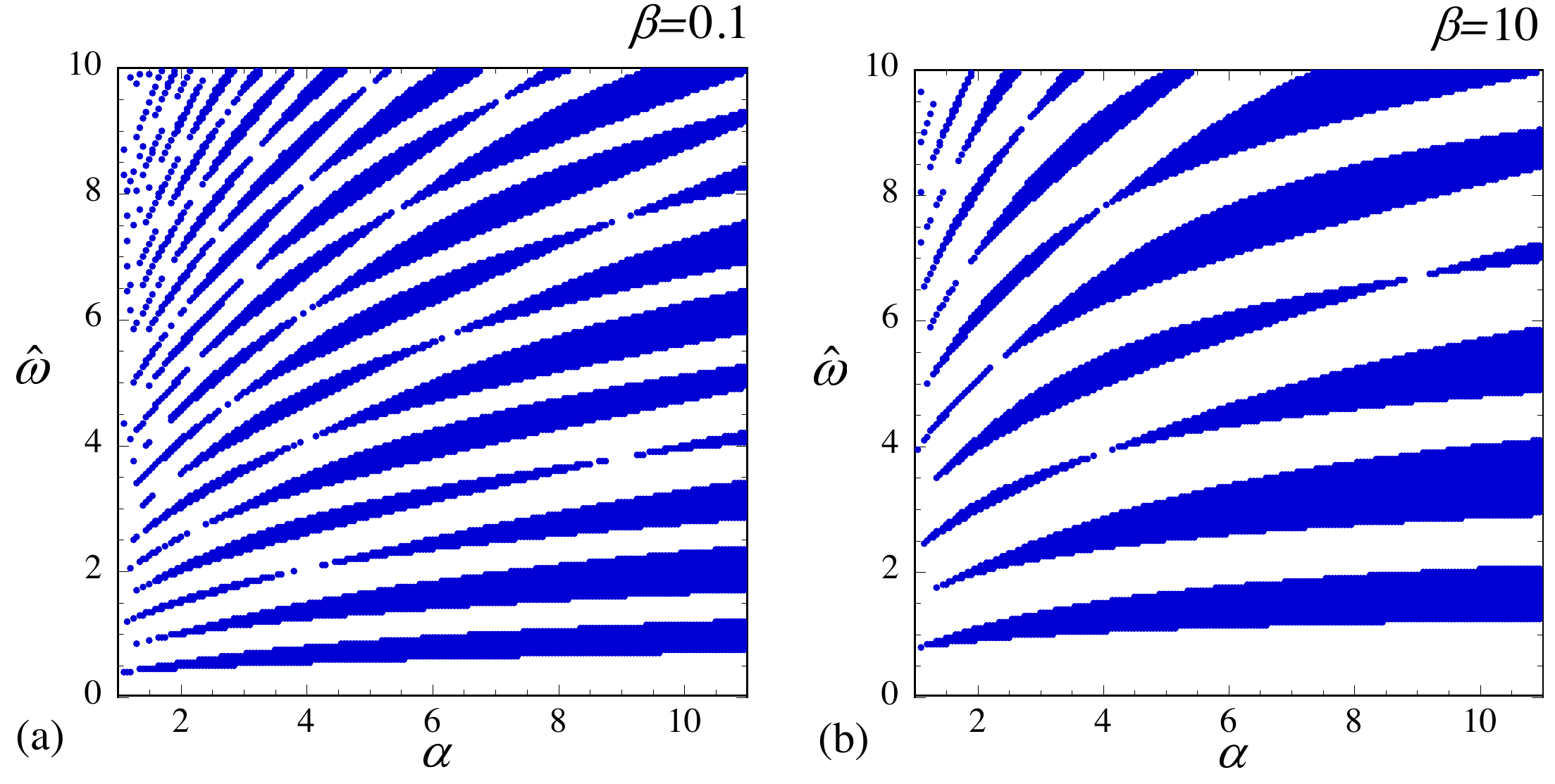} 
\caption{Prohibited frequencies as functions of $\alpha$ at $\dh=3$ and
$\mt v=0.5$, for (a) $\beta=0.1$ and (b) $\beta=10$.}
\label{bg of alpha}
\end{figure}

\begin{figure}[t]
\includegraphics[scale=0.45]{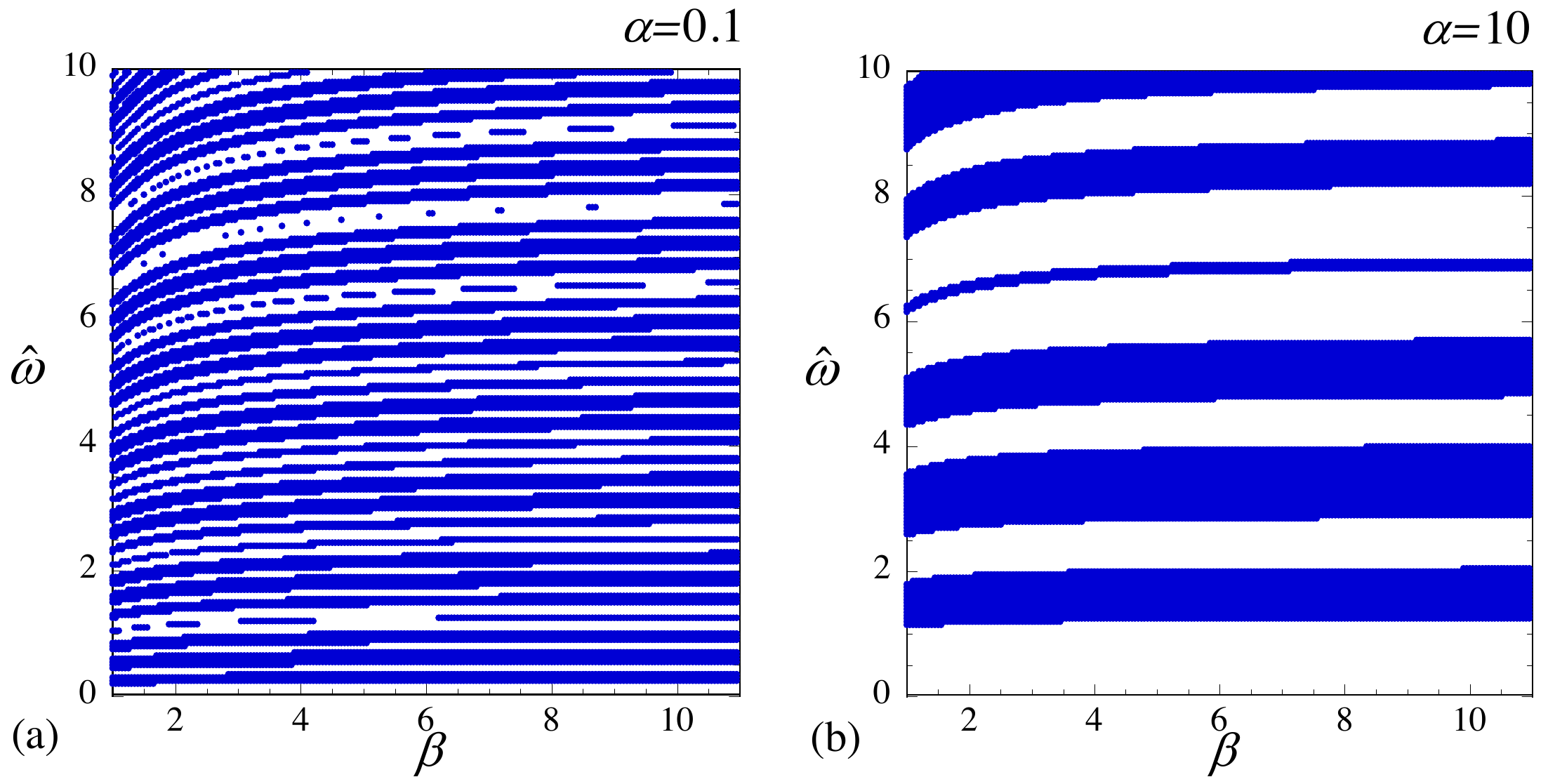}
\caption{Prohibited frequencies as functions of $\beta$ at $\dh=3$ and $\mt v=0.5$,
for (a) $\alpha=0.1$ and (b) $\alpha=10$.}
\label{bg of b}
\end{figure}

Fig.~\ref{bg of vm} displays the regions of prohibited normalized frequencies $\hat{\omega}$ as functions of the matrix volume fraction at a fixed value of the bias field $\dh=3$. 
Figs.~\ref{bg of vm}a, b and c correspond to $\alpha=\beta=5$, $10$ and $20$, respectively. 
Herein and henceforth
the band-gaps are the blue regions in the plots. 
We observe how an increase of the softer phase volume fraction results in appearance of additional
thinner bands. 
Conversely, when the contrasts are increased,
that is higher values of $\alpha$ and $\beta$, there are fewer thicker bands. 
The latter is evident in Fig.~\ref{bg of alpha} which
displays the prohibited frequencies as functions of $\alpha$ for $\mt v=0.5$
and $\dh=3$, with different values of $\beta$.
Specifically, Fig.~\ref{bg of alpha}a corresponds to $\beta=0.1$ and Fig.~\ref{bg of alpha}b
corresponds to $\beta=10$. 
Similar trends are observed when the dielectric
contrast is investigated. Fig.~\ref{bg of b} displays the prohibited
frequencies $\hat{\omega}$ as functions of $\beta$ for $\mt v=0.5$,
with different values of $\alpha$.
Fig.~\ref{bg of b}a corresponds to $\alpha=0.1$ and Fig.~\ref{bg of b}b to $\alpha=10$.
Figs.~\ref{bg of alpha} and \ref{bg of b} reveal that whenever the phases have non-monotonous ratios of shear moduli and dielectric contrasts,
i.e.~$\alpha>1$ and $\beta>1$ or $\alpha<1$ and $\beta>1$, the band gaps become thiner and denser.


\begin{figure}[t]
\includegraphics[angle=0,scale=0.60]{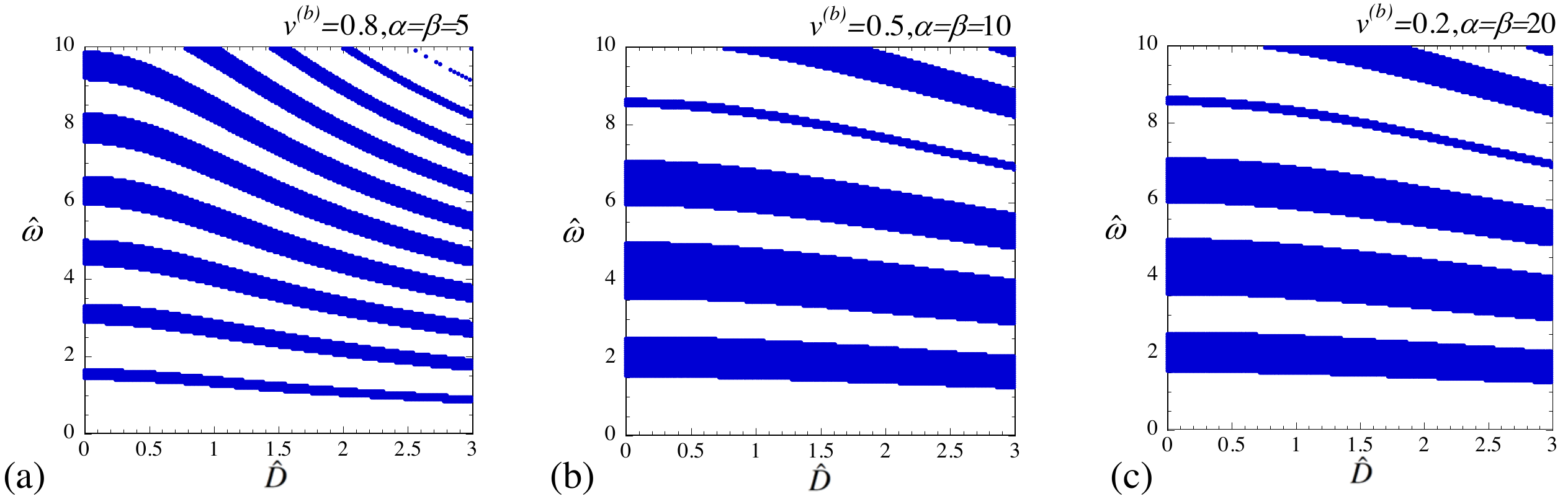}
\caption{Prohibited frequencies as functions of $\dh$ for (a) $\mt v=0.8,\alpha=\beta=5$, (b) $\mt v=0.5,\alpha=\beta=10$ and (c) $\mt v=0.2,\alpha=\beta=20$.}
\label{bg of d }
\end{figure}

Fig.~\ref{bg of d } illustrates the influence of the bias field by
examining the variations of the BGs as functions of the normalized
parameter $\dh$. Values of $\mt v=0.8$ and $\alpha=\beta=5$ were
chosen in Fig.~\ref{bg of d }a, $\mt v=0.5$ and $\alpha=\beta=10$
in Fig.~\ref{bg of d }b, and $\mt v=0.2$ and $\alpha=\beta=20$
in Fig.~\ref{bg of d }c. 
The influence of the electric field is evident as the BGs are shifted toward lower frequencies when the
field is increased. 
This phenomena suggests that \emph{control of
BGs and filtering of desired frequencies are feasible by activating
DE laminates with adequate bias electric fields.}

Finally, we consider realizable laminates made out of commercially available DEs. 
We examine three materials: VHB-4910 by 3M, ELASTOSIL
RT-625 by Wacker, and fluorosilicone 730 by Dow Corning. 
Approximate physical properties of these elastomers, as reported in the literature \citep[e.g., ][]{kofo&somm05sap,korn&pelr08inbook}, are summarized in table 1. 
 We investigate different combinations of the two-phase laminate and
assume that $\mt v=\fb v$. 

\begin{table}[b]
\caption{Physical properties of commercially available DEs. }
\begin{tabular}{|c|c|c|c|}
\hline 
{Material}  & density $\left[\rm{kg}/{m^{3}}\right]$ & shear modulus $\left[\rm kPa\right]$ & {relative dielectric}\tabularnewline
 & & & constant\\
\hline 
VHB-4910 & 960 & 406 & 4.7\tabularnewline
\hline 
ELASTOSIL RT-625  & 1020 & 342 & 2.7\tabularnewline
\hline 
Fluorosilicone 730 & 1400 & 167 & 6.9\tabularnewline
\hline 
\end{tabular}
\end{table}

\begin{figure}[t]
\includegraphics[scale=0.45]{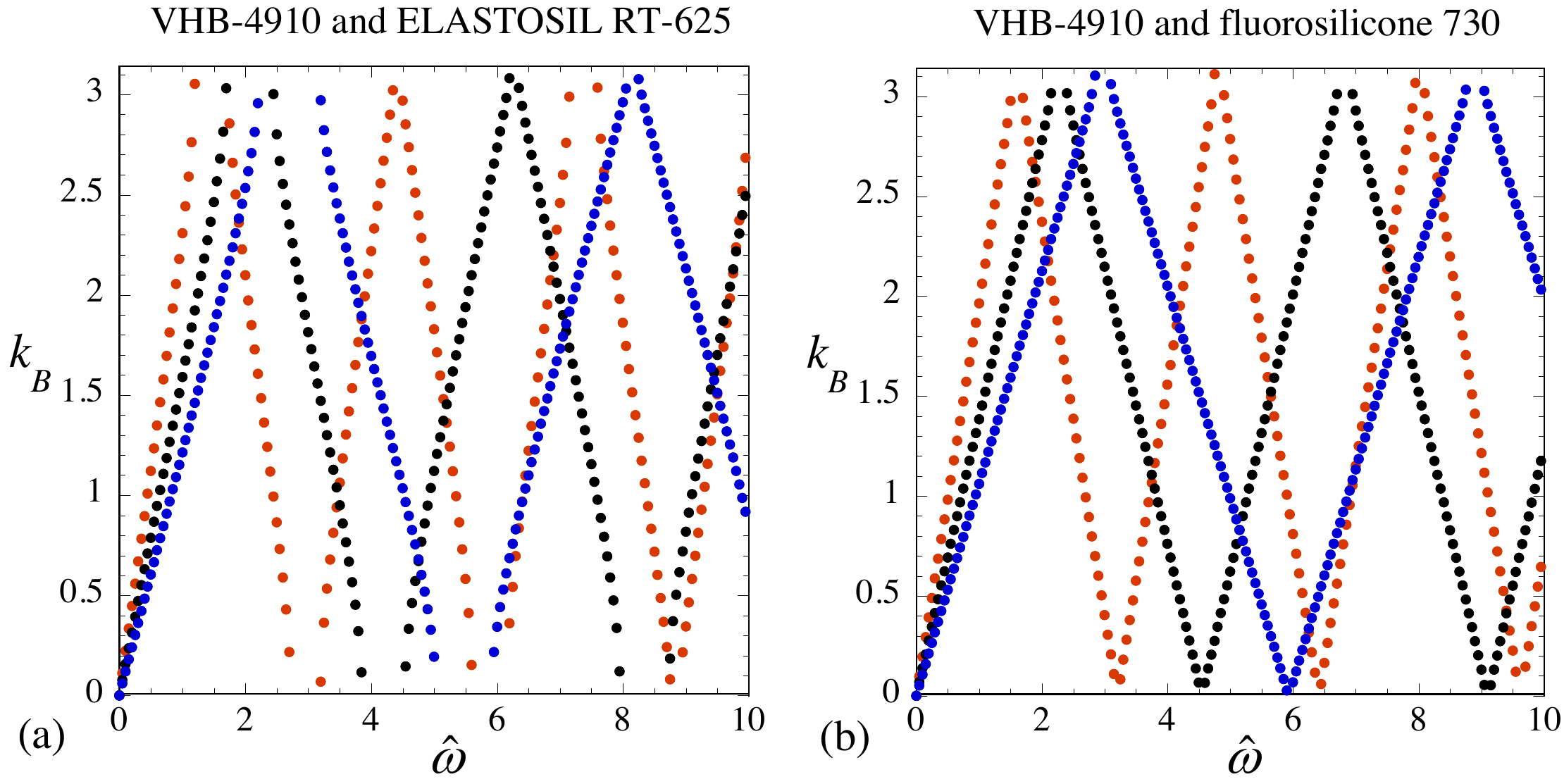}
\caption{Variations of the Bloch wavenumber $\kb$ as functions of the normalized frequency
$\afreqh=\afreq H/\cb$ for combinations of VHB-4910 with (a) ELASTOSIL
RT-625, and with (b) fluorosilicone 730. }
\label{dispersion}
\end{figure}

\begin{figure}[t]
\includegraphics[scale=0.45]{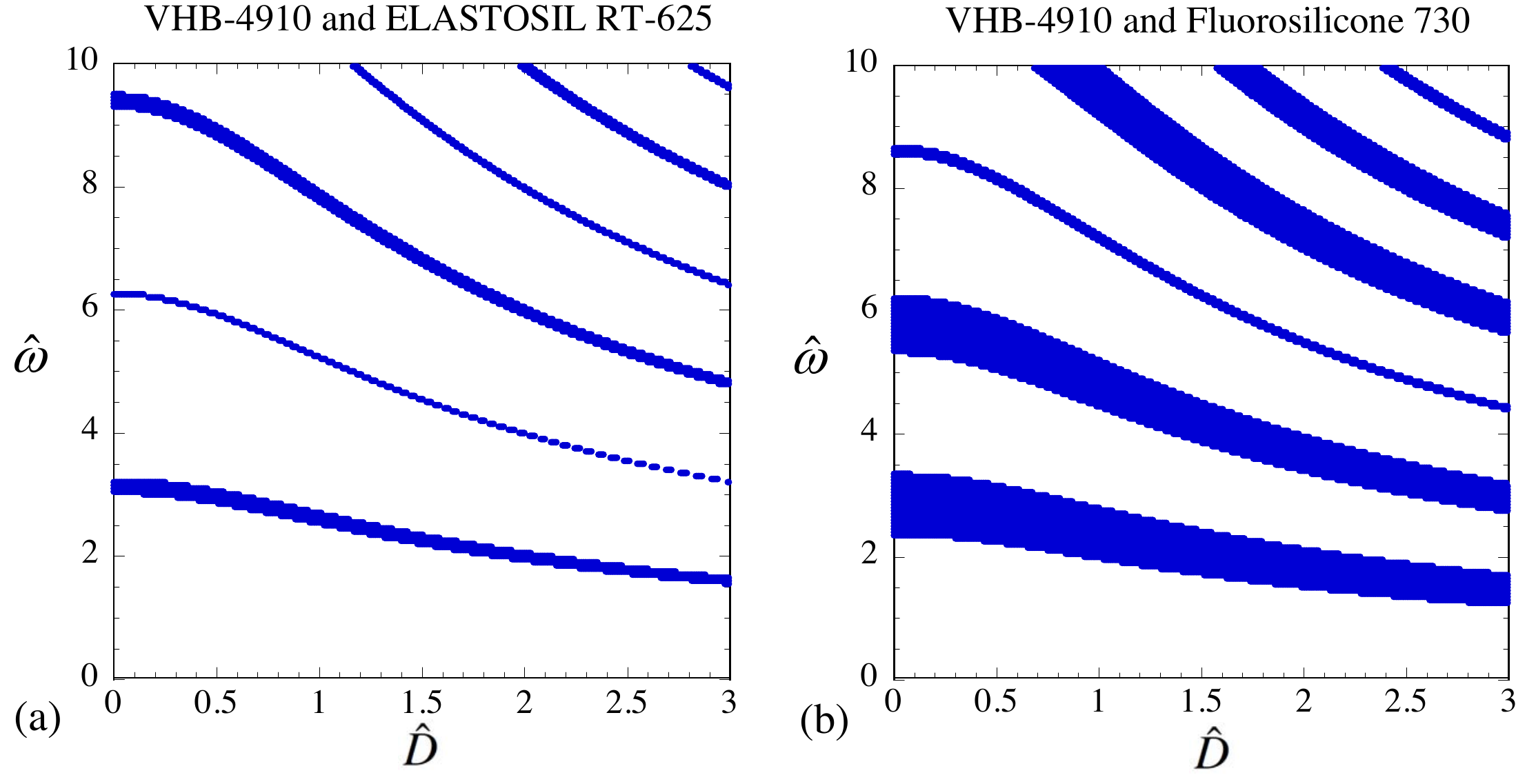} 
\caption{Prohibited frequencies as functions of $\dh$ for combinations of VHB-4910
with (a) ELASTOSIL RT-625, and (b) fluorosilicone 730. }
\label{bg}
\end{figure}

Fig.~\ref{dispersion} displays the dispersion relation in terms of
Bloch wavenumber $\kb$ as a function of the normalized frequency
$\afreqh$. Fig.~\ref{dispersion}a corresponds to a combination of
VHB-4910 and ELASTOSIL RT-625, while Fig.~\ref{dispersion}b to VHB-4910
and fluorosilicone 730. In terms of the dimensionless parameters in
Fig.~\ref{dispersion}a $\alpha=1.19$ and $\beta=1.74$, and in Fig.
\ref{dispersion}b $\alpha=2.44$ and $\beta=0.68$. The blue, black
and red curves correspond to $\dh=0.5,1$ and 1.5, respectively.

Fig.~\ref{bg} shows the prohibited frequencies as functions of the
bias field in terms of the dimensionless bias field $\dh$. Again,
Fig.~\ref{bg}a corresponds to the combination of VHB-4910 and ELASTOSIL
RT-625, and Fig.\ref{bg}b c to VHB-4910 and fluorosilicone 730. Wider
bands are depicted in Fig.~\ref{bg}b in comparison with Fig.~\ref{bg}a,
since the latter corresponds to a laminate with higher shear contrast
between the phases.

\section{Concluding remarks}

Motivated by the ability of DEs to undergo large deformations
and change their mechanical and electrical properties when electrostatically excited, the feasibility of inducing and controlling BGs by the electrical bias field is explored.
To this end we first considered the static response of an infinite periodic laminate to a bias electric displacement field. 
Subsequently, the propagation of small waves superposed on the pre-strained composite
is addressed. 
Applications of the transfer matrix method along with the
Bloch-Floquet theorem resulted in a dispersion relation between
the waves frequencies, the phase velocities, and the Bloch-parameter.
Numerical analysis of these relations was conducted for specific two-phase laminates in order to examine how the stop-bands, associated with complex Bloch-parameter, vary as functions of the geometrical, mechanical and electric properties of the laminate.
Most importantly, this analysis reveals the influence of the electrostatic bias fields on the BGs.
Our findings demonstrate how when the concentration of the softer phase in increased, the bands
become thinner and additional band gaps the appear.
An inverse effect is observed when the contrast between the phases properties is increased (larger values of $\alpha$ and $\beta$). 
The primary conclusion of our analysis is depicted in Fig.~\ref{bg of d }, demonstrating how
variations in the bias electric displacement field lead to modifications in the width and shifts in the range of the prohibited frequencies.
We note that for lower values of $\alpha$ and $\beta$ the effect of the
bias electric field becomes more evident in the sense that the shift in
the BGs range is larger. 
From practical viewpoint point, while stopping larger bands of frequencies is feasible by choosing a higher
contrast, lower contrasts will allow to actively adjust the desired band at a higher precision. 
The analysis was concluded with two examples in which we chose phases properties to be identical to
those of commercially available DEs.
The results show how tunable stop-bands are achievable by properly adjusting the electrostatic
bias field. 

\appendix

\section{The transfer matrix}

The components of the transfer matrices $\mathrm{T}^{\left(p\right)}_{(n)}$ appearing
in section 3:
\begin{align}
&\p{T_{11}}= \p{T_{22}}=\p{T_{33}}=\p{T_{55}}=\cos\p k\p h,\\
&\p{T_{13}}= \p{T_{52}}=\sin\p k\p h/\mb\p k,\\
&\p{T_{31}}= \p{T_{13}}=\p{T_{52}}=-\mb\p k\sin\p k\p h,\\
&\p{T_{66}}= -\p h/d\p{\matper},\\
&\p{T_{44}}= \p{T_{66}}=1,
&\end{align}
with all the remaining components being zero.

\begin{small}


\end{small}

\end{document}